\documentclass[twocolumn,preprintnumbers,amsmath,amssymb,prd,showkeys]{revtex4}

\usepackage{color, graphicx}
\usepackage{hyperref}

\begin{document}

\preprint{HIP-2007-48/TH}
\preprint{arXiv:0709.1319 [hep-th]}

\title{A Thermodynamic Interpretation of Time for Superstring Rolling Tachyons}

\author{Jimmy A. Hutasoit}
\email{jhutasoi@andrew.cmu.edu}
\affiliation{Carnegie Mellon University, Department of Physics, Pittsburgh, PA 15217, USA}
\author{Niko Jokela}
\email{niko.jokela@helsinki.fi}
\affiliation{Helsinki Institute of Physics, P.O. Box 64, FIN-00014, University of Helsinki, Finland}

\begin{abstract}
Rolling tachyon backgrounds, arising from open strings on unstable branes in bosonic string theory, can be related to a simple statistical mechanical model - Coulomb gas of point charges in two dimensions confined to a circle, the Dyson gas. In this letter we describe a statistical system that is dual to non-BPS branes in superstring theory. We argue that even though the concept of time is absent in the statistical dual sitting at equilibrium, the notion of time can emerge at the large number of particles $N \rightarrow \infty$ limit.
\end{abstract}

% \PACS code \sep code
 
\maketitle

\section{Introduction}
The behaviors of quantum fields in curved spacetime give hints that, as thermodynamic quantities, spacetime might be an emergent entity, emerging as an effective description of a system that does not contain it. String theory gives further motivations for this idea (for a review see \cite{Seiberg:2006wf}). Not only that, there have been a number of examples in which several dimensions of space emerge \cite{Klebanov:1991qa, Taylor:2001vb, Aharony:1999ti}.

What about the emergence of time? In a theory of quantum gravity ``time'' should also be an emergent concept -- there should be an underlying formalism which does not explicitly include a timelike coordinate. This suggests that the fundamental system is a Eucledian or a statistical system. As a step toward understanding the emergence of time, we can consider a (heuristic) analogy with AdS/CFT correspondence. If we think of AdS/CFT as emergence of space from a configuration of branes that are localized in space, then a candidate setting for investigating emergence of time will be branes that are localized in \emph{time} \cite{Gutperle:2002ai}. This is provided by the unstable branes of bosonic and superstring theory \cite{Sen:2002nu, Sen:2002in, Larsen:2002wc} (for a fairly recent review, see \cite{Sen:2004nf}). See also the initial proposal by Sen, where emergent time is viewed as the tachyon field that arises in the effective field theory description of the rolling tachyon \cite{Sen:2002qa}.

The emergence of time in the bosonic D-branes was investigated in \cite{Balasubramanian:2006sg}\footnote{For the so-called full S-brane -- a process of a creation and a subsequent decay of an unstable D-brane, see \cite{Jokela}.}, where the worldsheet description of the brane decay is related to a sequence of points of thermodynamic equilibrium of a grand canonical ensemble of point charges on a circle, the Dyson gas. There, subsequent instants of time are related to neighboring values of the average particle number $\bar{N}$, and the free energy of the system decreases as $\bar{N}$ increases (corresponding to later times), thus defining a thermodynamic arrow of time. In this letter, we will extend that work into the case of non-BPS branes in superstring theory and find analogous results. %See \cite{Balasubramanian:2006sg} for additional discussion.

\section{The Decaying Brane as a Paired Dyson Gas}
\subsection{Partition Function of the Decaying Brane}
Non-BPS brane of Type II superstring can be described as the exactly marginal deformation
\begin{eqnarray}
\delta S_{\rm bdry} = - \sqrt{2} \pi \lambda \int \frac{dt}{2 \pi} \psi_0 \, e^{X^0/\sqrt{2}}\otimes \sigma_1 \label{eq:deform} \ ,
\end{eqnarray}
where $\psi^0$ is the time component of the worldsheet fermion and $\sigma_1$ is a Chan-Paton factor associated with the boundary tachyon. The worldsheet correlation functions in this background then take the form
\begin{eqnarray}
\bar{A}_n &=& \int DX^0 \, DX^1 \cdots DX^9  D\psi^0 \, D\psi^1 \cdots D\psi^9 \nonumber \\
& & D\tilde{\psi}^0 \, D\tilde{\psi}^1 \cdots D\tilde{\psi}^9 \, e^{- S} \prod_{a = 1}^{n} V_a (z_a, \bar{z}_a) \label{eq:correlation} \ ,
\end{eqnarray}
where the action $S$ includes the boundary deformation (\ref{eq:deform}) and the $V_a$ are vertex operators. We can adopt convenient gauge choices \cite{Lambert:2003zr, Hwang:1991an}, where the dependence on the time component $X^0$ of the bosonic field takes a simple form
\begin{eqnarray}
V_a = e^{i \omega_a X^0} \, V_a^{\perp}(X^i, \psi^i, \tilde{\psi}^i, \cdots)
\end{eqnarray}
in the NS-NS sector, and
\begin{eqnarray}
V_a = e^{i \omega_a X^0} \, \Theta_{s_0} \tilde{\Theta}_{\tilde{s}_0} \, V_a^{\perp}(X^i, \psi^i, \tilde{\psi}^i, \cdots) \ ,
\end{eqnarray}
in the R-R sector. Here, the spin fields $\Theta_{s_0} = e^{i s_0 H^0}$ are in the bosonized form. The virtue of this gauge choice is that we can omit the trivial spatial part of the calculation of the correlations functions and just focus on the temporal part. The worldsheet correlation function (\ref{eq:correlation}) can be evaluated by isolating the zero mode $x^0$ from the fluctuations as $X^0 = x^0 + X'^0$, and expanding the boundary perturbation $e^{- \delta S_{\rm bdry}}$ in power series. The results can be written in a form
\begin{eqnarray}
\bar{A}_n = \int dx^0 e^{i x^0 \sum_a \omega_a} A_n(x^0) \ ,
\end{eqnarray}
where $A_n(x^0)$ can be written as a infinite power series of expectation values in Circular Unitary Ensembles \cite{Jokela:2005ha} (see also \cite{Shelton:2004ij}).

Let us first focus on the disk partition function $Z_{\rm open} = A_0 (x^0)$. The result reads {}\footnote{Here we are following the conventions of \cite{Balasubramanian:2004fz, Polchinski:1998rr} with $\alpha' = 1$.}
\begin{eqnarray}
 Z_{\rm open} & = & \sum_N (-)^N \frac{(\pi \lambda e^{x^0/\sqrt{2}})^{2N}}{N! \, N!} \nonumber\\
 & & \cdot\left(\int_{-\pi}^{\pi} \prod_{i = 1}^{N} \frac{dt_i}{2 \pi} \prod_{i<j} |e^{it_i} - e^{it_j}|^2 \right)^2 \ . \label{eq:diskcano}
\end{eqnarray}
Noticing that the integrand is the Haar measure for integration over $U(N)$ matrices, giving $N!$, we then have the following closed form
\begin{equation}\label{eq:opendisk}
Z_{\rm open} = \frac{1}{1 + \pi^2 \lambda^2 e^{\sqrt{2} \, x^0}} \ .
\end{equation}

\subsection{Partition Function of the Paired Dyson Gas}
Now we are ready to describe the statistical mechanical system that is dual to the non-BPS brane. Consider a gas of charged particles with infinitely heavy masses, confined to live on a unit circle on a two-dimensional plane, the Dyson gas. Pairs of these charges, having positions $e^{i t_j}$, interact through a two-dimensional Coulomb potential, given by
\begin{equation}
V^{\rm Dyson}(t_i, t_j) = - Q^2 \log|e^{i t_i} - e^{i t_j}| \ ,
\end{equation}
where $Q$ is the charge of the particles.

Consider now a system consisting of two separate subsystems: $N_+$ particles with charges $Q_+ = 1$ living on a boundary of a two-dimensional disk denoted by $v$, and $N_-$ particles with charges $Q_- = - 1$ living on a boundary of a different two-dimensional disk denoted by $w$, with $N_+ = N_- \equiv N$. Furthermore, particles living on different disks do not interact. One can imagine preparing the system with a perfect insulator wall separating the particles into two subsystems.

\begin{figure}[ht]
\begin{center}
\noindent
\includegraphics[width=0.45\textwidth]{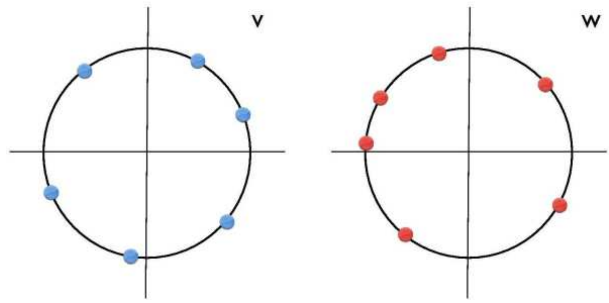}
\end{center}
\caption{Statistical dual of non-BPS brane: blue and red dots represent charges with $Q_+ = 1$ and $Q_- = - 1$, respectively.}
\label{fig:noinsertion}
\end{figure}

Suppose that this system is immersed into a heat bath at inverse temperature $\beta=1/T$. Now we aim to make contact with the disk partition function, by moving to consider the grand canonical ensemble with a particle reservoir with chemical potentials $\mu_+,\mu_-$. The grand canonical partition function of this system will then be
\begin{eqnarray}
Z_G &=& \sum_{N_+ = N_-} \left[\frac{{z_+}^{N_+}}{N_+!} \int_0^{2\pi} \prod_{i = 1}^{N_+} \frac{dt_i}{2\pi} \, e^{\left(- \beta \sum_{i<j} V^{\rm Dyson}(t_i,t_j) \right)} \right] \nonumber \\
& & \left[\frac{{z_-}^{N_-}}{N_-!} \int_0^{2\pi} \prod_{i = 1}^{N_-} \frac{d\tau_i}{2\pi} \, e^{\left(- \beta \sum_{k<l} V^{\rm Dyson}(\tau_k,\tau_l) \right)} \right] \label{eq:dysoncano} \ ,
\end{eqnarray}
where, as before, $V^{\rm Dyson}(x,y) = - \log |e^{ix} - e^{iy}|$. $N_\pm!$ take into account that the system consists of identical particles of two types, and $z_\pm = e^{\beta \mu_\pm}$ are the corresponding fugacities. At inverse temperature $\beta=2$ we then obtain
\begin{equation}\label{eq:dysonpart}
Z_G = \frac{1}{1 - z_+ z_-} \ .
\end{equation}

Comparing (\ref{eq:dysoncano}) with the open string disk partition function calculation (\ref{eq:diskcano}), by setting
\begin{equation}
z_+ = -z_- = \pi \lambda e^{x^0/\sqrt 2} \ ,
\end{equation}
one finds
\begin{equation}
 Z_G = \frac{1}{1+\pi^2\lambda^2 e^{\sqrt 2 x^0}} = Z_{\rm open} \ .
\end{equation} 
However, we would like to avoid having complex chemical potentials. If we want to end up with positive fugacity $z_-$, a simple remedy is to insert a $(-)^N$ into the sum (\ref{eq:diskcano}), and consider, instead of $Z_{\rm open}$, the quantity $Z'_{\rm open} \equiv \langle i^{\hat{N}_T} \rangle_{\rm brane}$. The virtue of this that it can be related to grand canonical partition function with positive fugacities. Another way of achieving this is to analytically continue $z_- \to -z_-$, while keeping $\lambda$ positive, so that we have one-to-one correspondence with time $x^0$ of the decaying brane and a positive fugacity,
\begin{equation}
 z \equiv |z_+| = |z_-| = \pi\lambda e^{x^0/\sqrt 2} \ .
\end{equation}
%See \cite{Balasubramanian:2006sg} for more extensive discussion on this.

\subsection{Correlation Functions}

To establish the relation between the non-BPS brane and the paired Dyson gas, we also need to prescribe how to calculate the worldsheet correlation functions in the Dyson gas picture. For clarity, in this subsection we will explicitly write $z_\pm$ in favor of $z$.

\subsubsection{NS-NS Vertex Operators}

The statistical dual will correspond to inserting a set of test charges with charges $q_+^a = i \omega_a /\sqrt{2}$ at $v = z_a$ and another set of test charges with charges $q_-^a = - i \omega_a /\sqrt{2}$ at $w = z_a$.

\begin{figure}[ht]
\begin{center}
\noindent
\includegraphics[width=0.45\textwidth]{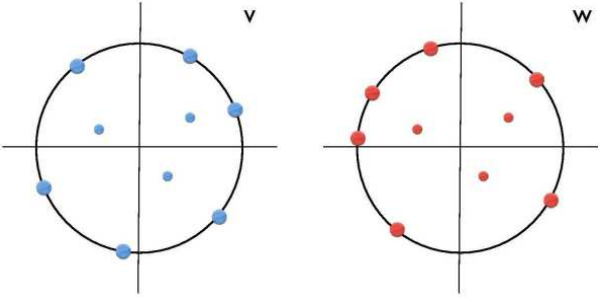}
\end{center}
\caption{Statistical dual of NS-NS vertex operators insertion: smaller blue and red dots represent test charges with $q_+^a = i \omega_a /\sqrt{2}$ at $v = z_a$, and $q_-^a = - i \omega_a /\sqrt{2}$ at $w = z_a$, respectively.}\label{fig:NS-NS}
\end{figure}

To be more precise, let us consider inserting vertex operators in the form of $V_s = \prod_{a = 1}^n e^{i \omega_a X^0(z_a, \bar{z}_a)}$. To compute the correlation functions of this kind in the Dyson gas picture, we need to compute the grand canonical ``expectation value''  of the quantity $e^{- \beta \, V^{\rm bulk}}$, where $V^{\rm bulk}$ is the interaction among bulk charges. This is defined as
\begin{eqnarray}
& &\langle \prod_{a = 1}^n e^{i \omega_a X^0(z_a, \bar{z}_a)} \rangle \equiv \nonumber \\
& &\sum_N \left(\frac{z_+^N}{N!} \int \prod_{i = 1}^{N} \frac{dt_i}{2 \pi} \, e^{\left[ -\beta \left(V_+^{\rm Dyson} + V_+^{\rm bulk - Dyson} \right) \right]} \cdot e^{- \beta \, V_+^{\rm bulk}} \right) \nonumber \\
& &\left(\frac{z_-^N}{N!} \int \prod_{i = 1}^{N} \frac{d\tau_i}{2 \pi} \, e^{\left[ -\beta \left(V_-^{\rm Dyson} + V_-^{\rm bulk - Dyson} \right) \right]} \cdot e^{- \beta \, V_-^{\rm bulk}} \right), \nonumber \\
\end{eqnarray}
where $V^{\rm bulk - Dyson}$ is the interaction between the bulk charges and the Dyson gas particles.

\subsubsection{R-R Vertex Operators}

For correlation functions involving R-R sector, we need to do two things:
\begin{itemize}
\item add $|k|$ Dyson particles with charges $Q = {\rm sign(k)}$ to the respective two-dimensional plane (\emph{i.e.}, $Q = + 1$ to $v$-plane, $Q = - 1$ to $w$-plane). Here $k$ is defined as $k \equiv - \sum_a (s_a + \tilde{s}_a) \in \mathbb{Z}$;
\item insert test charges with $q_+^a = (i \omega_a /\sqrt{2} + (s_a + \tilde{s}_a)/2)$ at $v = z_a$, and $q_-^a = (- i \omega_a /\sqrt{2} + (s_a + \tilde{s}_a)/2)$ at $w = z_a$, respectively.
\end{itemize}

\begin{figure}[ht]
\begin{center}
\noindent
\includegraphics[width=0.45\textwidth]{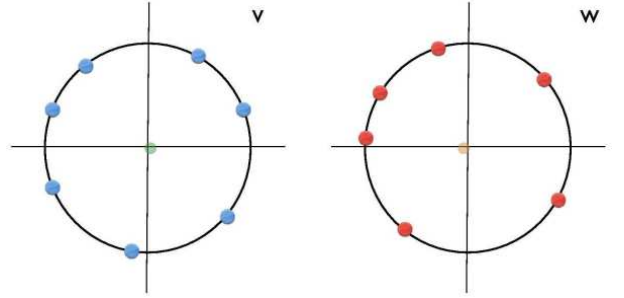}
\end{center}
\caption{Statistical dual of R-R vertex operators insertion: green and orange dots represent test charges with $q_+^a = (i \omega_a /\sqrt{2} + (s_a + \tilde{s}_a)/2)$ at $v = z_a$, and $q_-^a = (- i \omega_a /\sqrt{2} + (s_a + \tilde{s}_a)/2)$ at $w = z_a$, respectively. Here, as an example, we have taken $s_a = \tilde{s}_a = 1/2$ so that $k = 1$. Thus, there is an extra $q = 1$ Dyson particle on the boundary of $v$-disk.}\label{fig:R-R}
\end{figure}

Let us consider inserting vertex operators in the form of $\prod_{a = 1}^n \Theta_{s_a}(z_a) \tilde{\Theta}_{\tilde{s}_a}(\bar{z}_a)$ $e^{i \omega_a X^0(z_a, \bar{z}_a)}$. The prescription for calculating the expectation value in the Dyson gas picture then becomes
\begin{eqnarray}
& &\langle \prod_{a = 1}^n \Theta_{s_a}(z_a) \tilde{\Theta}_{\tilde{s}_a}(\bar{z}_a) e^{i \omega_a X^0(z_a, \bar{z}_a)} \rangle \equiv \nonumber \\
& & \sum_N \left(\frac{z_+^{N + |k|}}{(N + |k|)!} \int \prod_{i = 1}^{N + |k|} \frac{dt_i}{2 \pi} \, e^{\left[ -\beta \left(V_+^{\rm Dyson} + V_+^{\rm bulk - Dyson} \right) \right]}  \right) \nonumber \\
& &\cdot \,\, e^{- \beta V_+^{\rm bulk}} \nonumber \\
& & \left(\frac{z_-^N}{N!} \int \prod_{i = 1}^{N} \frac{d\tau_i}{2 \pi} \, e^{\left[ -\beta \left(V_-^{\rm Dyson} + V_-^{\rm bulk - Dyson} \right) \right]} \cdot e^{- \beta V_-^{\rm bulk}} \right) \nonumber \\
\end{eqnarray}
for $k \geq 0$, or
\begin{eqnarray}
& & \langle \prod_{a = 1}^n \Theta_{s_a}(z_a) \tilde{\Theta}_{\tilde{s}_a}(\bar{z}_a) e^{i \omega_a X^0(z_a, \bar{z}_a)} \rangle \equiv \nonumber \\
& & \sum_N  \left(\frac{z_-^{N + |k|}}{(N + |k|)!} \int \prod_{i = 1}^{N + |k|} \frac{d\tau_i}{2 \pi} \, e^{\left[ -\beta \left(V_-^{\rm Dyson} + V_-^{\rm bulk - Dyson} \right) \right]} \right) \nonumber \\
& &  \cdot \,\, e^{- \beta V_-^{\rm bulk}} \nonumber \\
& & \left(\frac{z_+^{N}}{N!} \int \prod_{i = 1}^{N} \frac{dt_i}{2 \pi} \, e^{\left[ -\beta \left(V_+^{\rm Dyson} + V_+^{\rm bulk - Dyson} \right) \right]} \cdot e^{- \beta V_+^{\rm bulk}} \right) \nonumber \\
\end{eqnarray}
for $k \leq 0$.

\section{The Emergence of Time}
In the Dyson gas picture, the time of the target spacetime $x^0$ has been replaced by the chemical potential $\mu$ of the statistical system. Different points of thermodynamic equilibrium characterized by different values of the chemical potential correspond to the different instants in time for the decaying non-BPS brane.

\subsection{Time as the Average Particle Number}
We can now give an interpretation to time as the average particle number in the statistical system. The average particle numbers for both sectors of the Dyson gases are
\begin{equation}
 \bar{N} = \frac{ z^2}{1 - z^2} = \frac{\pi^2 \lambda^2 e^{\sqrt{2} x^0}}{1 - \pi^2 \lambda^2 e^{\sqrt{2} x^0}}.
\end{equation}
Solving for $x^0$ we get
\begin{equation}
x^0 = \frac{1}{\sqrt{2}} \log \frac{\bar{N}}{1 + \bar{N}} - \sqrt{2} \log \pi \lambda \ .
\end{equation}

At the past infinity, $x^0 \rightarrow - \infty$, we have vanishing fugacity and average particle number. Then $\bar{N}$ increases monotonically as a function of the fugacity corresponding to later time values of $x^0$.

$\bar{N}$, like time, is a continuous quantity. However, the underlying physical quantity is $N$, which is discrete. $N$ fluctuates around $\bar{N}$, and the fluctuation is given by
\begin{equation}
\frac{\delta N}{\bar{N}} = \frac {\sqrt{\langle N^2 \rangle - \langle N \rangle^2}}{\bar{N}} = \frac{1}{z}  = \frac{e^{- x^0/\sqrt{2}}}{\pi \lambda} \ .
\end{equation}
At early times, $\delta N$ is large, but later $\bar{N}$ becomes more sharply defined. We can interpret this as a continuous time emerging from an underlying discrete variable $N$ in the large $N$ limit.

\subsection{Thermodynamic Arrow of Time}
Let us consider the grand potential of the statistical system
\begin{equation}
\Omega (z, T, V) = - T \log Z_G.
\end{equation}
Using Legendre transformation, we can get the Helmholtz free energy
\begin{eqnarray}
A(\bar{N}, T, V) = - \frac{1}{2} \left[\log (\bar{N} + 1)^{(\bar{N} + 1)} - \log \bar{N}^{\bar{N}} \right].
\end{eqnarray}
We can rewrite $A = - T I_S$ where $I_S$ is the Shannon entropy. Shannon entropy in the particle number distribution is given by
\begin{eqnarray}
I_S = - \sum_n p(n) \log p(n); \,\,\,\,\, p(n) = \frac{(z^2)^n}{Z_G},
\end{eqnarray}
where $p(n)$ is the probability that there are $n$ positively charged particles on $v$-disk and $n$ negatively charged particles on $v$-disk.  Since $I_s$ increases as $\bar{N}$ increases, the free energy decrease as $\bar{N}$ increases. Recalling that the increase of $\bar{N}$ marks the passage of time, we can interpret this as a thermodynamic arrow of time.

\section{Discussion}
In this paper, we have reformulated the worldsheet description of non-BPS brane decay in terms of the statistical mechanics of the paired Dyson gas. As in the case of bosonic D-brane decay, the progress of time was marked by the increase of the average number of particles in the gas. The decrease of free energy marked the passage of time and thus gave rise to a thermodynamic arrow of time.

We would like to emphasize that even though one might have expected a non-equilibrium flow to realize the flow of time, this expectation is somewhat misleading. This is because there is already a physical time in non-equilibrium statistical mechanics. Thus, one can not expect to have time emerging from a non-equilibrium statistical mechanical system. Instead, in this set-up, we found that each instant of time was described by a different point of equilibrium in a statistical system.

In the known examples of emergence spatial dimensions, we learned that the classical or geometrical picture of spacetime is only valid in a certain limit. Here we can also see that it is true, in the sense that the classical geometry of time, \emph{i.e.}, $\mathbb{R}$ emerges in the limit of large number of particles. Furthermore, we would also like to emphasize that this classical geometry of time does not correspond to a thermal state of the paired Dyson gas, but corresponds to the \emph{whole} ensemble of the gas (\emph{c.f.}, black hole does not correspond to the thermal state but the whole ensemble of its microstates \cite{Balasubramanian:2007qv}).

If the notion of spacetime being emergent is true, then it is a possibility that spacetime symmetries are emergent too. In our case, the spacetime picture of the homogeneously decaying unstable D-brane is believed to be supersymmetric at late times, while on the other hand the Dyson gas is not. Then, one could ask whether the spacetime supersymmetry must be emergent too and how this is realized in Dyson gas language. This certainly deserves more study.

% This might be merely a consequence of the statistical dual corresponding to $A_n(x^0)$ instead of corresponding to $\tilde{A}_n(x^0, % h^0)$, defined as
% \begin{equation}
% \bar{A}_n = \int dh^0 dx^0 e^{i x^0 \sum_a \omega_a} \tilde{A}_n(x^0, h^0),
% \end{equation}
% where $h^0$ is the fermionic zero mode.

\vspace{.3cm}
{\bf{Acknowledgements}}
\vspace{.3cm}\newline
The authors would like to thank Richard Holman, Esko Keski-Vakkuri, Jaydeep Majumder, Jessie Shelton, and Robert Swendsen for useful comments and discussions. N.J. has been in part supported by the Magnus Ehrnrooth foundation. This work was also partially supported by the EU 6th Framework Marie Curie Research and Training network ``UniverseNet'' (MRTN-CT-2006-035863). J.H. was supported in part by DOE grant DE-FG03-91-ER40682. N.J. thanks the Carnegie Mellon University and University of Illinois at Urbana-Champaign for hospitality while this work was in progress.


\begin{thebibliography}{00}
%\cite{Seiberg:2006wf}
\bibitem{Seiberg:2006wf}
  N.~Seiberg,
  ``Emergent spacetime,''
  \href{http://www.slac.stanford.edu/spires/find/hep/www?irn=6507905}{arXiv:hep-th/0601234}.
  %%CITATION = HEP-TH/0601234;%%
%\cite{Klebanov:1991qa}
\bibitem{Klebanov:1991qa}
  I.~R.~Klebanov,
  ``String Theory In Two-Dimensions,''
  \href{http://www.slac.stanford.edu/spires/find/hep/www?r=pupt-1271}{arXiv:hep-th/9108019}.
  %%CITATION = HEP-TH/9108019;%%
%\cite{Taylor:2001vb}
\bibitem{Taylor:2001vb}
  W.~Taylor,
  ``M(atrix) theory: Matrix quantum mechanics as a fundamental theory,''
  Rev.\ Mod.\ Phys.\  {\bf 73} (2001) 419
  \href{http://www.slac.stanford.edu/spires/find/hep/www?j=rmpha\%2c73\%2c419}{[arXiv:hep-th/0101126]}.
  %%CITATION = RMPHA,73,419;%%
%\cite{Aharony:1999ti}
\bibitem{Aharony:1999ti}
  O.~Aharony, S.~S.~Gubser, J.~M.~Maldacena, H.~Ooguri and Y.~Oz,
  ``Large N field theories, string theory and gravity,''
  Phys.\ Rept.\  {\bf 323} (2000) 183
  \href{http://www.slac.stanford.edu/spires/find/hep/www?j=prplc\%2c323\%2c183}{[arXiv:hep-th/9905111]}.
  %%CITATION = PRPLC,323,183;%%
%\cite{Gutperle:2002ai}
\bibitem{Gutperle:2002ai}
  M.~Gutperle and A.~Strominger,
  ``Spacelike branes,''
  JHEP {\bf 0204} (2002) 018
  \href{http://www.slac.stanford.edu/spires/find/hep/www?j=jhepa\%2c0204\%2c018}{[arXiv:hep-th/0202210]}.
  %%CITATION = JHEPA,0204,018;%%
%\cite{Sen:2002nu}
\bibitem{Sen:2002nu}
  A.~Sen,
  ``Rolling tachyon,''
  JHEP {\bf 0204} (2002) 048
 \href{http://www.slac.stanford.edu/spires/find/hep/www?j=jhepa\%2c0204\%2c048}{[arXiv:hep-th/0203211]}.
  %%CITATION = JHEPA,0204,048;%%
%\cite{Sen:2002in}
\bibitem{Sen:2002in}
  A.~Sen,
  ``Tachyon matter,''
  JHEP {\bf 0207} (2002) 065
  \href{http://www.slac.stanford.edu/spires/find/hep/www?j=jhepa\%2c0207\%2c065}{[arXiv:hep-th/0203265]}.
  %%CITATION = JHEPA,0207,065;%%
%\cite{Larsen:2002wc}
\bibitem{Larsen:2002wc}
  F.~Larsen, A.~Naqvi and S.~Terashima,
  ``Rolling tachyons and decaying branes,''
  JHEP {\bf 0302} (2003) 039
  \href{http://www.slac.stanford.edu/spires/find/hep/www?j=jhepa\%2c0302\%2c039}{[arXiv:hep-th/0212248]}.
  %%CITATION = JHEPA,0302,039;%%
%\cite{Sen:2004nf}
\bibitem{Sen:2004nf}
  A.~Sen,
  ``Tachyon dynamics in open string theory,''
  Int.\ J.\ Mod.\ Phys.\  A {\bf 20} (2005) 5513
  \href{http://www.slac.stanford.edu/spires/find/hep/www?j=impae\%2ca20\%2c5513}{[arXiv:hep-th/0410103]}.
  %%CITATION = IMPAE,A20,5513;%%
%\cite{Sen:2002qa}
\bibitem{Sen:2002qa}
  A.~Sen,
  ``Time and tachyon,''
  Int.\ J.\ Mod.\ Phys.\  A {\bf 18} (2003) 4869
  \href{http://www.slac.stanford.edu/spires/find/hep/www?j=impae\%2ca18\%2c4869}{[arXiv:hep-th/0209122]}.
  %%CITATION = IMPAE,A18,4869;%%
%\cite{Balasubramanian:2006sg}
\bibitem{Balasubramanian:2006sg}
  V.~Balasubramanian, N.~Jokela, E.~Keski-Vakkuri and J.~Majumder,
  ``A thermodynamic interpretation of time for rolling tachyons,''
  Phys.\ Rev.\  D {\bf 75} (2007) 063515
  \href{http://www.slac.stanford.edu/spires/find/hep/www?j=phrva\%2cd75\%2c063515}{[arXiv:hep-th/0612090]}.
  %%CITATION = PHRVA,D75,063515;%%
%\cite{Jokela}
\bibitem{Jokela}
  N.~Jokela, E.~Keski-Vakkuri and J.~Majumder,
  ``Timelike Boundary Sine-Gordon Theory and Two-Component Plasma,''
  \href{http://www.slac.stanford.edu/spires/find/hep/www?r=hip-2007-47-th}{arXiv:0709.1318 [hep-th]}.
  %%CITATION = ARXIV:0709.1318;%%
%\cite{Lambert:2003zr}
\bibitem{Lambert:2003zr}
  N.~Lambert, H.~Liu and J.~M.~Maldacena,
  ``Closed strings from decaying D-branes,''
  JHEP {\bf 0703} (2007) 014
  \href{http://www.slac.stanford.edu/spires/find/hep/www?j=jhepa\%2c0703\%2c014}{[arXiv:hep-th/0303139]}.
  %%CITATION = JHEPA,0703,014;%%
%\cite{Hwang:1991an}
\bibitem{Hwang:1991an}
  S.~Hwang,
  ``Cosets As Gauge Slices In SU(1,1) Strings,''
  Phys.\ Lett.\  B {\bf 276} (1992) 451
  \href{http://www.slac.stanford.edu/spires/find/hep/www?j=phlta\%2cb276\%2c451}{[arXiv:hep-th/9110039]}.
  %%CITATION = PHLTA,B276,451;%%
%\cite{Jokela:2005ha}
\bibitem{Jokela:2005ha}
  N.~Jokela, E.~Keski-Vakkuri and J.~Majumder,
  ``On superstring disk amplitudes in a rolling tachyon background,''
  Phys.\ Rev.\  D {\bf 73} (2006) 046007
  \href{http://www.slac.stanford.edu/spires/find/hep/www?j=phrva\%2cd73\%2c046007}{[arXiv:hep-th/0510205]}.
  %%CITATION = PHRVA,D73,046007;%%
%cite{Shelton:2004ij}
\bibitem{Shelton:2004ij}
  J.~Shelton,
  ``Closed superstring emission from rolling tachyon backgrounds,''
  JHEP {\bf 0501} (2005) 037
  \href{http://www.slac.stanford.edu/spires/find/hep/www?j=jhepa\%2c0501\%2c037}{[arXiv:hep-th/0411040]}.
  %%CITATION = JHEPA,0501,037;%%
%\cite{Balasubramanian:2004fz}
\bibitem{Balasubramanian:2004fz}
  V.~Balasubramanian, E.~Keski-Vakkuri, P.~Kraus and A.~Naqvi,
  ``String scattering from decaying branes,''
  Commun.\ Math.\ Phys.\  {\bf 257} (2005) 363
  \href{http://www.slac.stanford.edu/spires/find/hep/www?j=cmpha\%2c257\%2c363}{[arXiv:hep-th/0404039]}.
  %%CITATION = CMPHA,257,363;%%
%\cite{Polchinski:1998rr}
\bibitem{Polchinski:1998rr}
  J.~Polchinski,
  ``String theory. Vol. 2: Superstring theory and beyond,''
%\href{http://www.slac.stanford.edu/spires/find/hep/www?irn=4634802}{SPIRES entry}
{\it  Cambridge, UK: Univ. Pr. (1998) 531 p}
%\cite{Balasubramanian:2007qv}
\bibitem{Balasubramanian:2007qv}
  V.~Balasubramanian, B.~Czech, V.~E.~Hubeny, K.~Larjo, M.~Rangamani and J.~Simon,
  ``Typicality versus thermality: An analytic distinction,''
  \href{http://www.slac.stanford.edu/spires/find/hep/www?r=upr-t-1170}{arXiv:hep-th/0701122}.
  %%CITATION = HEP-TH/0701122;%%
\end{thebibliography}
\end{document}